\documentclass{JHEP3}

  \usepackage{latexsym,bm,amsmath,amssymb,amsfonts}
  \usepackage{epsfig,graphics,graphicx}
  \usepackage{slashed}
  \usepackage[latin1]{inputenc}
  \usepackage{mathrsfs}
\usepackage{mathcomp}

  \long\def\comment#1{ }

\def\be{\begin{equation}}
\def\ee{\end{equation}}
\def\bea{\begin{eqnarray}}
\def\eea{\end{eqnarray}}
\def\nn{\nonumber}

\title{\rm \LARGE \bf Remarks on self-interaction correction to black hole radiation}

\author{Qing-Quan Jiang$^{a,b}$ and Xu Cai$^{b}$\\

$^{a}$ Institute of Theoretical Physics, China
 West Normal University, Nanchong, Sichuan \\637002, People's Republic of
China\\
$^{b}$ Institute of Particle Physics, Central China Normal
University, Wuhan, Hubei 430079, \\People's Republic of China\\
{\tt E-mail address: jiangqq@iopp.ccnu.edu.cn, xcai@mail.ccnu.edu.cn}}

\abstract{In the work [P. Kraus and F. Wilczek, \textit{Self-interaction
correction to black hole radiation, Nucl. Phys.} B433 (1995) 403],
it has been pointed out that the self-gravitation interaction
would modify the black hole radiation so that it is no longer
thermal, where it is, however, corrected in an approximate way and
therefore is not established its relationship with the underlying
unitary theory in quantum theory. In this paper, we revisit the
self-gravitation interaction to Hawking radiation of the general
spherically symmetric black hole, and find that the precisely
derived spectrum is not only deviated from the purely thermal
spectrum, but most importantly, is related to the change of the
Bekenstein-Hawking entropy and consistent with an underlying
unitary theory.}

\begin{document}

%%%%%%%%%%%%%%%%%%%%%%%%%%%%%%%%%%%%%%%%%%%%%%%%%%%%%%%%%%%%%
\section{Introduction}
%%%%%%%%%%%%%%%%%%%%%%%%%%%%%%%%%%%%%%%%%%%%%%%%%%%%%%%%%%%%%
In 1974, Hawking proved that black hole can radiate particles from
its event horizon with a temperature proportional to its surface
gravity, and the radiant spectrum is a purely thermal one
\cite{r1}, where the background geometry was approximately given
by calculating the response of quantum fields to the collapse
geometry. Now it is difficult to reconcile with the unitary
evolution in quantum theory, which finally generates an
information loss of black hole. In the classical theory, the loss
of information was not a serious problem since the information
could be thought of as preserved inside the black hole but just
not very accessible. However, taking the quantum effect into
consideration, the situation is changed. With the emission of
thermal radiation, black holes could lose energy, shrink, and
eventually evaporate away completely. Since the radiation with a
precise thermal spectrum carries no information, the information
carried by a physical system falling toward black hole singularity
has no way to be recovered after a black hole has disappeared
completely. This is the so-called ``information loss paradox",
which means that pure quantum states (the original matter that
forms the black hole) can evolve into mixed states (the thermal
spectrum at infinity)\cite{r2}. Such an evolution violates the
fundamental principles of quantum theory, as these prescribe a
unitary time evolution of basis states. To resolve the information
loss paradox, one must go beyond the approximation of treating the
background geometry as fixed, and treat it as a quantum variable
and then quantize the gravitational system.

In 1995, Kraus and Wilczek found the black hole radiation is no
longer thermal after considering its self-gravitation interaction
\cite{r3}. Its idea is based on the Hamiltonian quantization,
where the coupled particle-hole system is treated as the quantized
system. After eliminating the non-physical degrees of freedom and
quantizing the reduced system, the derived radiation deviates from
the purely thermal spectrum. This elaborate work is invaluable due
to the facts that: 1) it may provide a possible explanation to the
problem of the information loss; 2) it can open a window to study
the quantum gravity. Till now, the two open questions are still
unsolved and present at the frontier of the modern theoretical
physics. So it is necessary to find some new physics by solving
the self-gravitational interaction on black hole radiation via the
Hamiltonian quantization method.

In \cite{r3}, it has been proved that the self-gravitational
interaction would modify the black hole radiation, but the way in
which it is corrected is not definite. In this paper, we revisit
the self-gravitation interaction to Hawking radiation of the
spherically symmetric black hole, and find that the precisely
derived spectrum is not only deviated from the purely thermal
spectrum, but most importantly, is related to the change of
Bekenstein-Hawking entropy and consistent with an underlying
unitary theory. This result presents a definite way in which the
black hole radiation is corrected as a result of its
self-gravitational interaction, and moreover guarantees the
unitary evolution in black hole quantum radiation.

The remainders of this paper are outlined as follows. In
Sec.\ref{sec1}, we first solve the constraint conditions to
eliminate the non-physical degrees of freedom for the
particle-hole system, and keep only those degrees of freedom which
are most relevant to the problem of particle emission from the
regions of low curvature. Sec.\ref{sec2} is to quantize the
reduced particle-hole system, and present the self-interaction
correction to black hole radiation. Sec.\ref{sec3} ends up with
some discussions and conclusions.

%%%%%%%%%%%%%%%%%%%%%%%%%%%%%%%%%%%%%%%%%%%%%%%%%%%%%%%%%%%%%
\section{Eliminating the non-physical degrees of freedom for the particle-hole system } \label{sec1}
%%%%%%%%%%%%%%%%%%%%%%%%%%%%%%%%%%%%%%%%%%%%%%%%%%%%%%%%%%%%%
The four dimensional metric in the
ADM(Arnowitt-Deser-Misner)(namely, (3+1) decomposition) form can
be written as \cite{RSC,LZ} \bea
ds^2&=&-\left(\mathcal{N}^2(t,x^i)-\mathcal{N}_i(t,x^i)\mathcal{N}^i(t,x^i)\right)dt^2\nn\\
&+&2\mathcal{N}_i(t, x^i)dx^idt+h_{ij}(t,x^i)dx^idx^j,
\label{eq1}\eea
 where $\mathcal{N}\equiv
(-g^{00})^{-1/2}$ is formally called as the time shift,
$\mathcal{N}_i\equiv g_{0i}$ as the space shift, and
$\mathcal{N}^i=h^{ij}\mathcal{N}_j$. This decomposition for the
gravitational field can be found in Appendix A.
 Now the new expression of the
action for the shell-hole system in the ADM form can be written as
 \bea
 S&=&\int
d^4x\sqrt{h}\mathcal{N}\left(\mathcal{R}^{(3)}+\mathcal{K}_{ij}\mathcal{K}^{ij}-\mathcal{K}^2\right)\nn\\
&-& m\int dt
\sqrt{\mathcal{\hat{N}}^2-\mathcal{\hat{N}}_i\mathcal{\hat{N}}^i-2\mathcal{\hat{N}}_i\dot{\hat{x}}^i-\hat{h}_{ij}
\dot{\hat{x}}^i\dot{\hat{x}}^j}+ 2\int
\sqrt{h}\mathcal{K}d^3x,\label{eq2} \eea where
$\mathcal{K}_{ij}=\frac{1}{2\mathcal{N}}\left[\dot{h}_{ij}-\left(\mathcal{N}_{i|j}+\mathcal{N}_{j|i}\right)\right]$,
in which $i|j$ is the covariant derivative on the metric $h_{ij}$,
and $\mathcal{K}=h^{ij}\mathcal{K}_{ij}$. In addition,
$\mathcal{R}^{(3)}$ denotes the scalar curvature of the
3-dimensional spacial metric $h_{ij}$. $m$ is the rest mass of the
shell, and the carets instruct one to evaluate physical quantities
at the shell (as an example,
$\mathcal{\hat{N}}=\mathcal{N}(\hat{t},\hat{x}^i)$ ). Obviously,
the independent variables for the action in the ADM form appear as
$h_{ij}$, $\mathcal{N}$ and $\mathcal{N}_i$. Since Hamiltonian
method is more flexible than Lagrangian method in eliminating
constraints, in this paper we resort to Hamiltonian method to
eliminate the non-physical degrees of freedom for the
particle-hole system. To do so, we must first write the action
(\ref{eq2}) in the canonical form. In our case, we consider the
the emitted particles coupled to the spherically symmetric
gravitational fields, and its reduced 3-dimensional spacial line
element is taken the form as
\begin{equation}
ds_3^2=e^{2\mu}
dr^2+e^{2\lambda}\left(d\theta^2+\sin\theta^2d\phi^2\right).\label{eq3}
 \end{equation}
Assuming $\pi^{ij}$ is the canonical momentum with respect to the
3-dimensional spacial metric $h_{ij}$, we have
\begin{equation}
\pi^{rr}=\frac{1}{2}\pi_\mu e^{-2\mu}, ~~~~
\pi^{\theta\theta}=\frac{1}{4}\pi_\lambda e^{-2\lambda}, ~~~~
\pi^{\phi\phi}=\frac{1}{4}\pi_\lambda e^{-2\lambda},
 \end{equation}
 where $\pi_\mu$ and $\pi_\lambda$ are the momentum conjugate to
 $\mu$ and $\lambda$. Now the action of the gravity-shell system in the canonical form
can be written as
 \bea
 S&=&\int dtdr \left[\pi_\mu
 \dot{\mu}+\pi_\lambda \dot{\lambda}
 -\mathcal{N}\left(\mathcal{H}_g^0+\mathcal{H}_s^0\right)-\mathcal{N}_r\left(\mathcal{H}_g^r+\mathcal{H}_s^r\right)\right]\nn\\
&+&\int dt p_r \dot{\hat{r}}-\int dt M_{ADM}, \label{eq5}
 \eea
where
\bea
 \mathcal{H}_s^0&=& \sqrt{p_r^2e^{-2\mu}+m^2}\delta(r-\hat{r}),
 ~~~~~~~~\mathcal{H}_s^r=-p_re^{-2\mu}\delta(r-\hat{r}),\nn\\
\mathcal{H}_g^0&=&e^{-\mu-2\lambda}\left[\frac{1}{2}\pi_\mu^2-\pi_\mu\pi_\lambda+\frac{1}{2}e^{4\lambda}\left(2\lambda''-2\lambda'\mu'+3\lambda'-e^{2(\mu-\lambda)}\right)\right],\nn\\
\mathcal{H}_g^r&=&
-e^{-2\mu}\left(\pi_\mu'-\mu'\pi_\mu-\lambda'\pi_\lambda\right).
\label{eq6}
 \eea
 Here the last term in Eq.(\ref{eq5}) is present to cancel
the unwanted terms when carrying on the variation of the action
\cite{r3}. The general relativity is a theory with constraints, so
we wish to eliminate the gravitational degrees of freedom to
obtain the effective action only depending on the shell variables
by solving the constraints. Since the action (\ref{eq5}) is not
the functional of the parameters $\dot{\mathcal{N}}$ and
$\dot{\mathcal{N}_r}$, the constraints are obtained by varying
with respect to $\mathcal{N}$ and $\mathcal{N}_r$ as
$\mathcal{H}^0=\mathcal{H}_g^0+\mathcal{H}_s^0=0$ (the energy
constraint) and $\mathcal{H}^r=\mathcal{H}_g^r+\mathcal{H}_s^r=0$
(the momentum constraint). Obviously, the momentum $\pi_\mu$ and
$\pi_\lambda$ can be obtained from the constraints. For
simplicity, we, here, consider the linear combination of
constraints as \bea
0&=&e^{\lambda-\mu}\lambda'\mathcal{H}^0+e^{-\lambda}\pi_\mu
\mathcal{H}^r\nn\\
& =&
-\frac{1}{2}\partial_r\big(e^{-2\mu-\lambda}\pi_\mu^2+e^\lambda-e^{\lambda-2\mu}(e^\lambda
\lambda')^2\big)\nn\\
&+&\hat{e}^{\lambda-\mu}\hat{\lambda}'\mathcal{H}_s^0+\hat{e}^{-\lambda}\hat{\pi}_\mu
\mathcal{H}_s^r.\label{eq7}
 \eea
We can obtain from the constraint (\ref{eq7}) that at the shell
$\hat{\pi}_\mu=0$ and $\hat{\lambda}'=0$, which implies
${\lambda}'$ is discontinuous here. Substituting the physical
quantities at the shell into the energy constraint equation and
integrating them, we have
$\lambda'(\hat{r}+\epsilon)-\lambda'(\hat{r}-\epsilon)=\hat{e}^{-2\lambda}\sqrt{p_r^2+m^2\hat{e}^{-2\mu}}$.
If we define $\mathcal{M}\equiv
\frac{1}{2}(e^{-2\mu-\lambda}\pi_\mu^2+e^\lambda-e^{\lambda-2\mu}(e^\lambda
\lambda')^2)$, it can be treated as the mass parameter when
considering a static slice ($\pi_\mu=\pi_\lambda=0$). Away from
the shell, the constraints tell us $\mathcal{M}$ is a constant. At
the shell, its presence causes $\mathcal{M}$ to be discontinuous.
Here we take the forms of the discontinuous mass parameter in
different regions as $\mathcal{M}=M$ at $(r<\hat{r})$, and $M=M_+$
at $(r>\hat{r})$. As there is no matter outside the shell, we have
$M_{ADM}=M_+$. Now the momentum $\pi_\mu$ can be written as \bea
\pi_\mu&=& e^{\mu+\lambda}\sqrt{\left(e^\lambda
\lambda'\right)^2e^{-2\mu}-1+2Me^{-\lambda}},~~~~~~~r<\hat{r}, \nn\\
\pi_\mu&=& e^{\mu+\lambda}\sqrt{\left(e^\lambda
\lambda'\right)^2e^{-2\mu}-1+2M_+e^{-\lambda}},~~~~~r>\hat{r}.
\label{eq8}
 \eea
When the constraints are satisfied, considering a variation of the
action yields $dS=p_r d{\hat{r}}+\int dr\left(\pi_\mu
\delta\mu+\pi_\lambda \delta\lambda\right)-M_+dt$. To obtain the
effective action, we must integrate the expression for an
arbitrary shell trajectory. In our case, when holding $\hat{r}$,
$p_r$ and $\lambda$ fixed, varying $\mu$ until
$\pi_\mu=\pi_\lambda=0$ for an arbitrary trajectory yields
 \bea
S&=&\int_{r_{min}}^\infty dr \int_0^\mu \pi_\mu \delta\mu \nn\\
&=& \int_{r_{min}}^{\hat{r}-\epsilon}dr
\Big[e^{\mu+\lambda}\sqrt{(e^\lambda
\lambda')^2e^{-2\mu}-1+2Me^{-\lambda}}\nn\\
&+&e^\lambda (e^\lambda \lambda')\log\Big|\frac{(e^\lambda
\lambda')e^{-\mu}-\sqrt{(e^\lambda
\lambda')^2e^{-2\mu}-1+2Me^{-\lambda}}}{\sqrt{|1-2Me^{-\lambda}|}}\Big|\Big]\nn\\
&+& \int_{\hat{r}+\epsilon}^{\infty}dr
\Big[e^{\mu+\lambda}\sqrt{(e^\lambda
\lambda')^2e^{-2\mu}-1+2M_+e^{-\lambda}}\nn\\
&+&e^\lambda (e^\lambda \lambda')\log\Big|\frac{(e^\lambda
\lambda')e^{-\mu}-\sqrt{(e^\lambda
\lambda')^2e^{-2\mu}-1+2M_+e^{-\lambda}}}{\sqrt{|1-2M_+e^{-\lambda}|}}\Big|\Big].\label{eq9}
 \eea
From Eq.(\ref{eq9}), we can easily learn the variables are now
turn to be $\mu$, $\lambda$, $\lambda'$ and $M_+$. In the next
stage, if an arbitrary variation of $\mu$ and $\lambda$ is carried
on the effective action (\ref{eq9}), it changes as
\bea
dS&=&\int_{r_{min}}^\infty dr\Big(\pi_\lambda
\delta \lambda+\pi_\mu \delta\mu\Big)\nn\\
& -&\Big[\frac{\partial S}{\partial
\hat{\lambda}'}(\hat{r}+\epsilon)-\frac{\partial S}{\partial
\hat{\lambda}'}(\hat{r}-\epsilon)\Big]d\hat{\lambda}'-\frac{\partial
S}{\partial M_+}d M_+,
\eea
 where the last two terms are present
to keep the relations $\frac{\delta S}{\delta
\lambda}=\pi_\lambda$ and $\frac{\delta S}{\delta \mu}=\pi_\mu$
unchanged when carrying on the arbitrary variations of $\lambda$
and $\mu$. Now we can additionally consider variations in the
variables $p_r$, $\hat{r}$ and $t$ to obtain the final effective
action, however it is no need to separately consider variations of
$p_r$ and $\hat{r}$ the constraint equations contain their
variations, since its contributions to the effective action has
already been included in the constraints. As for the time variable
$t$, its contribution to the action is $dS=-M_+dt$. Finally, the
action can be written as \bea S &=&
\int_{r_{min}}^{\hat{r}-\epsilon}dr
\Big[e^{\mu+\lambda}\sqrt{(e^\lambda
\lambda')^2e^{-2\mu}-1+2Me^{-\lambda}}\nn\\
&+&e^\lambda (e^\lambda \lambda')\log\Big|\frac{(e^\lambda
\lambda')e^{-\mu}-\sqrt{(e^\lambda
\lambda')^2e^{-2\mu}-1+2Me^{-\lambda}}}{\sqrt{|1-2Me^{-\lambda}|}}\Big|\Big]\nn\\
&+& \int_{\hat{r}+\epsilon}^{\infty}dr
\Big[e^{\mu+\lambda}\sqrt{(e^\lambda
\lambda')^2e^{-2\mu}-1+2M_+e^{-\lambda}}\nn\\
&+&e^\lambda (e^\lambda \lambda')\log\Big|\frac{(e^\lambda
\lambda')e^{-\mu}-\sqrt{(e^\lambda
\lambda')^2e^{-2\mu}-1+2M_+e^{-\lambda}}}{\sqrt{|1-2M_+e^{-\lambda}|}}\Big|\Big]\nn\\
&-& \int dt
\dot{\hat{\lambda}}\hat{e}^{2\lambda}\Big[\log\Big|\frac{\Big(\hat{e}^\lambda
\lambda'(\hat{r}-\epsilon)\Big)\hat{e}^{-\mu}-\sqrt{\Big(\hat{e}^\lambda
\lambda'(\hat{r}-\epsilon)\Big)^2\hat{e}^{-2\mu}-1+2M\hat{e}^{-\lambda}}}{\sqrt{|1-2M\hat{e}^{-\lambda}|}}\Big|\nn\\
&-&\log\Big|\frac{\Big(\hat{e}^\lambda
\lambda'(\hat{r}+\epsilon)\Big)\hat{e}^{-\mu}-\sqrt{\Big(\hat{e}^\lambda
\lambda'(\hat{r}+\epsilon)\Big)^2\hat{e}^{-2\mu}-1+2M_+\hat{e}^{-\lambda}}}{\sqrt{|1-2M_+\hat{e}^{-\lambda}|}}\Big|\Big]\nn\\
&-&\int dt \int_{\hat{r}+\epsilon}^\infty dr
\frac{e^\mu}{\sqrt{(e^\lambda
\lambda')^2e^{-2\mu}-1+2M_+e^{-\lambda}}}\dot{M}_+-\int dt M_+.
\label{eq11}
 \eea

Differentiating (\ref{eq11}), we can successfully recover the
variation of the action where the constraints are satisfied. This
shows (\ref{eq11}) is the correct expression of the action. At the
shell, $\lambda'$ is discontinuous, and its form arbitrarily near
the shell is constrained by
$\lambda'(\hat{r}+\epsilon)-\lambda'(\hat{r}-\epsilon)=\hat{e}^{-2\lambda}\sqrt{p_r^2+m^2\hat{e}^{-2\mu}}$.
Near the shell but far enough away so that $\lambda'$ can be well
specifiable, we can define $\lambda_f'=\lambda'$. Arbitrarily near
the horizon, we can well define
$\lambda_f'=\lambda'(\hat{r}+\epsilon)$, and
$\lambda'(\hat{r}-\epsilon)$ can then be determined by the
discontinuous constraint condition. The Lagrangian function is
defined by the derivative of the action $S$ on the time $t$, which
yields
 \bea \mathcal{L}&=&\frac{dS}{dt} =
\dot{\hat{r}}\hat{e}^{\mu+\lambda}\Big[\sqrt{(\hat{e}^\lambda
\lambda_f')^2\hat{e}^{-2\mu}-1+2M\hat{e}^{-\lambda}}\nn\\
&-&\sqrt{(\hat{e}^\lambda
\lambda_f')^2\hat{e}^{-2\mu}-1+2M_+\hat{e}^{-\lambda}}\Big]\nn\\
&-&
\dot{\hat{\lambda}}\hat{e}^{2\lambda}\log\Big|\frac{\big(\hat{e}^\lambda
\lambda'(\hat{r}-\epsilon)\big)\hat{e}^{-\mu}-\sqrt{\big(\hat{e}^\lambda
\lambda'(\hat{r}-\epsilon)\big)^2\hat{e}^{-2\mu}-1+2M\hat{e}^{-\lambda}}}{\big(\hat{e}^\lambda
\lambda_f'\big)\hat{e}^{-\mu}-\sqrt{\big(\hat{e}^\lambda
\lambda_f'\big)^2\hat{e}^{-2\mu}-1+2M_+\hat{e}^{-\lambda}}}\Big|\nn\\
&+&\int_{min}^{\hat{r}-\epsilon}dr\big(\pi_\lambda
\dot{\lambda}+\pi_\mu
\dot{\mu}\big)+\int_{\hat{r}+\epsilon}^{\infty}dr\big(\pi_\lambda
\dot{\lambda}+\pi_\mu \dot{\mu}\big)-M_+.\label{eq12}
 \eea
 Considering the discontinuous constraint condition of $\lambda'$ arbitrarily near the shell,
 and the shell mass approaching to infinite small, the Lagrangian
 function (\ref{eq12}) becomes
\bea \mathcal{L}&=&\frac{dS}{dt} =
\dot{\hat{r}}\hat{e}^{\mu+\lambda}\Big[\sqrt{(\hat{e}^\lambda
\lambda_f')^2\hat{e}^{-2\mu}-1+2M\hat{e}^{-\lambda}}\nn\\
&-&\sqrt{(\hat{e}^\lambda
\lambda_f')^2\hat{e}^{-2\mu}-1+2M_+\hat{e}^{-\lambda}}\Big]\nn\\
&-&\eta
\dot{\hat{\lambda}}\hat{e}^{2\lambda}\log\Big|\frac{\big(\hat{e}^\lambda
\lambda_f'\big)\hat{e}^{-\mu}-\eta\sqrt{\big(\hat{e}^\lambda
\lambda_f'\big)^2\hat{e}^{-2\mu}-1+2M_+\hat{e}^{-\lambda}}}{\big(\hat{e}^\lambda
\lambda_f'\big)\hat{e}^{-\mu}-\eta\sqrt{\big(\hat{e}^\lambda
\lambda_f'\big)^2\hat{e}^{-2\mu}-1+2M\hat{e}^{-\lambda}}}\Big|\nn\\
&+&\int_{min}^{\hat{r}-\epsilon}dr\big(\pi_\lambda
\dot{\lambda}+\pi_\mu
\dot{\mu}\big)+\int_{\hat{r}+\epsilon}^{\infty}dr\big(\pi_\lambda
\dot{\lambda}+\pi_\mu \dot{\mu}\big)-M_+,\label{rr1}
 \eea
 where $\eta\equiv\pm$ corresponds to the momentum of the outgoing (ingoing)
particle. Since at the black hole horizon, the particle radiates
out, we should choose the $+$ sign in the following discussion.
  Now
 the Lagrangian function has already been decided. To quantize the reduced
 particle-hole system we must first find the canonical coordinate
 (namely the physical degrees of freedom) and its corresponding canonical
 momentum. So we here must choose a gauge to specifically determine the form of
 the Lagrangian function (\ref{rr1}). In this paper, we take the
 Schwarzschild black hole as an example to show in what definite way
 Hawking radiation is corrected due to self-gravitation
 interaction. The metric for a general spherically symmetric system in ADM form
 has been shown in Eqs.(\ref{eq1}) and (\ref{eq3}). For a four-dimensional spherically
 Schwarzschild solution $\mathcal{N}=1$,
 $\mathcal{N}_i=\sqrt{\frac{2M}{r}}$, $e^\mu=1$ and $e^\lambda=r$.
 The Schwarzschild black hole in ADM form has been present in Appendix B.
At the shell, $\hat{e}^\lambda\lambda'$ is discontinuous, but in
Eq.(\ref{rr1}), $\hat{e}^\lambda \lambda'_f$ is still freely
specifiable. For the Schwarzschild black hole, the spacetime is
static (namely, time independent), so the Lagrangian can be
written as
\begin{equation}
\mathcal{L}=\dot{\hat{r}}\big(\sqrt{2M\hat{r}}-\sqrt{2M_+\hat{r}}\big)-\dot{\hat{r}}
\hat{r}\log\frac{\sqrt{\hat{r}}-\sqrt{2M_+}}{\sqrt{\hat{r}}-\sqrt{2M}}-M_+.\label{eq13}
\end{equation}
From Eq.(\ref{eq13}), the canonical momentum conjugate to
$\hat{r}$ is given by
\begin{equation}
p_{\hat{r}}=\frac{\partial\mathcal{L}}{\partial\dot{\hat{r}}}=\sqrt{2M\hat{r}}-\sqrt{2M_+\hat{r}}-
\hat{r}\log\frac{\sqrt{\hat{r}}-\sqrt{2M_+}}{\sqrt{\hat{r}}-\sqrt{2M}},\label{eq14}
\end{equation}
which means the action for the particle-hole system in the
canonical form can be written as
\begin{equation}
S=\int dt \mathcal{L}=\int dt
\big(p_{\hat{r}}\dot{\hat{r}}+p_t\big), \label{eq15}
\end{equation}
where $p_t=M-M_+$ comes from the contribution of the shell, and
denotes the canonical momentum conjugate to the time $t$. Here we
have omitted the term $\int dt M$, Since as a constant its
contribution to the variation of the effective action can not
change the physical result we want. Till now, we have eliminated
the non-physical degrees of freedom for the particle-hole system,
and obtained the effective action which only depended on the shell
variables. In the next section, we focus on the quantization of
the effective action and the self-interaction correction to
Hawking radiation.

%%%%%%%%%%%%%%%%%%%%%%%%%%%%%%%%%%%%%%%%%%%%%%%%%%%%%%%%%%%%%
\section{Self-interaction correction to Hawking radiation} \label{sec2}
%%%%%%%%%%%%%%%%%%%%%%%%%%%%%%%%%%%%%%%%%%%%%%%%%%%%%%%%%%%%%
To pass to the quantum theory we should make the substitutions
$p\rightarrow-i\frac{\partial}{\partial r}$ and $p_t \rightarrow
-i\frac{\partial}{\partial t}$. But when returning to our case,
how to implement this substitutions mets with factor ordering
ambiguities. Fortunately, the mode solutions $\nu_k(t, r)$ to the
field equations which are accurately described by the WKB
approximation, are insensitive to these difficulties. Near the
horizon, the infinite redshift results in a very short wavelength
for the modes involved, so the WKB approximation is valid here.
Therefore we can write these solutions as
$\nu_k(t,r)=e^{iS(t,r)}$. Introducing the replacements
$p_{\hat{r}}\rightarrow \frac{\partial S}{\partial \hat{r}}$ and
$p_t\rightarrow \frac{\partial S}{\partial t}$, we can obtain the
Hamilton-Jacobi equation for $S$. The solution of the  the
Hamilton-Jacobi equation is just the classical action. So, if
$\hat{r}(t)$ is a classical trajectory derived by extremizing
(\ref{eq15}), we have
\begin{equation}
S(t, \hat{r}(t))=S(0,\hat{r}(0))+\int_0^t dt
\big(p_{\hat{r}}(\hat{r}(t)) \dot{\hat{r}}(t)+p_t\big).
\label{eq16}
\end{equation}
In Eq.(\ref{eq16}), as the initial condition
$S(0,\hat{r}(0))\equiv k\hat{r}(0)$, where $k$ must be arbitrarily
large due to the ever increasing redshift experienced by the
emitted quanta as they escape to infinity. As an arbitrary
redshift existing at the horizon, when substituting (\ref{eq14})
into (\ref{eq16}) we can only keep the terms which is singular at
the horizon. Thus we have
\begin{equation}
S(t,r)=k\hat{r}(0)-\int_{\hat{r}(0)}^rd\hat{r}\hat{r}
\log\frac{\sqrt{\hat{r}}-\sqrt{2M_+}}{\sqrt{\hat{r}}-\sqrt{2M}}+(M-M_+)t,\label{eq17}
\end{equation}
where
\begin{equation}
k=\frac{\partial S}{\partial r}(0, \hat{r}(0))=-\hat{r}(0)
\log\frac{\sqrt{\hat{r}(0)}-\sqrt{2M_+}}{\sqrt{\hat{r}(0)}-\sqrt{2M}}.\label{eq18}
\end{equation}
For the Schwarzschild black hole, the black hole horizon is
coincident with the infinite redshift surface. So at the horizon,
$k\rightarrow \infty$, which yields from Eq.(\ref{eq18})
\begin{equation}
\hat{r}(0)=2M_+\pm \epsilon,\label{eq19}
\end{equation}
where $\epsilon$ is a small constant, ``$+$'' corresponds to
$M_+=M+\omega_k$, and ``$-$'' to $M_+=M-\omega_k$. Next, we focus
on studying the emission rate of the black hole, including effects
due to self-interaction. Black hole radiance originates from the
mismatch between the two natural vacuum states which arise in the
quantization of a field propagating on a black hole space-time.
The one is the vacuum state which is natural from the standpoint
of an observer making measurements at the infinity, and the other
vacuum state to consider is the one naturally employed by an
observer freely falling through the horizon. As for the first
vacuum state, we consider the modes which are positive frequency
with respect to the Killing vector. We write these modes as
$\mu_k(r)e^{-i\omega_kt}$. Thus the field operator can be expanded
in these modes as
\begin{equation}
{\phi}=\int
dk\big({a}_k\mu_k(r)e^{-i\omega_kt}+{a}_k^\dagger\mu_k^\ast(r)e^{i\omega_kt}\big).\label{eq20}
\end{equation}
Here the freely falling observer would find an infinite energy
momentum density, since the modes $\mu_k(r)$ are singular at the
horizon. In this case, the vacuum state is defined by
$a_k|0_\mu\rangle
 = 0$. To well describe the physical quantities at the horizon,
 this vacuum state is not expected, and the other vacuum state which is
 well behaved at the horizon is required. Writing the complete
 set of such modes by $\nu_k(t,r)$, the field operator can also be
 expanded as
\begin{equation}
{\phi}=\int
dk\big({b}_k\nu_k(t,r)+{b}_k^\dagger\nu_k^\ast(t,r)\big),\label{eq21}
\end{equation}
where the vacuum state is determined by ${b}_k|0_\nu\rangle=0$.
This vacuum state $|0_\nu\rangle$ is well behaved at the horizon,
so it can result in a finite energy-momentum density here. The
operators in Eqs.(\ref{eq20}) and (\ref{eq21}) are related by the
Bogoliubov coefficients as $ a_k=\int
dk'\big({\alpha}_{kk'}{b}_k+{\beta}_{kk'}b_{k'}^\ast\big)$, where
\bea
\alpha_{kk'}&=&\frac{1}{2\pi\mu_k(r)}\int_{-\infty}^{\infty}dt
e^{i\omega_kt}\nu_{k'}(t,r),\nn\\
\beta_{kk'}&=&\frac{1}{2\pi\mu_k(r)}\int_{-\infty}^{\infty}dt
e^{i\omega_kt}\nu_{k'}^\ast(t,r). \label{eq22}
 \eea
These coefficients in (\ref{eq22}) can be obtained in the saddle
point approximation. As $\nu_k=e^{iS(t,r)}$, the saddle point for
$\alpha_{kk'}$ is given by $\omega_k=-\partial_tS(t,r)=M_+-M$.
From Eq.(\ref{eq19}), this condition produces the integration
region of the effective action $S$ to lie in between
$\hat{r}(0)=2(M+\omega_k)+ \epsilon$ and outside the place
$\hat{r}(0)$. After integrating Eq.(\ref{eq17}) in this region, we
will find the action appears purely real, which yields
$|\alpha_{kk'}|^2=\frac{1}{|2\pi\mu_k(r)|^2}$ in the saddle
approximation.  For $\beta_{kk'}$, the saddle point is determined
by $\omega_k=\partial_tS(t,r)=M-M_+$, which present an imaginary
part for the action. In fact, if we assume that the mass of the
Schwarzschild black hole is held fixed whereas the total ADM mass
of the hole-particle system are allowed to fluctuate, when
radiating a particle with the energy ``$-\omega_k$'' from the
black hole horizon, we also obtain the ADM energy
$M_+=M-\omega_k$. At this point, we have
\begin{equation}
|\beta_{kk'}|^2=\frac{1}{|2\pi\mu_k(r)|^2}\exp{\Big(2\omega_k
\textrm{Im} t+2\textrm{Im}[S^\ast_k(t)]\Big)}.
\end{equation}
The emission rate is given by the effective Boltzmann factor as
\bea
\Big|\frac{\beta_{kk'}}{\alpha_{kk'}}\Big|^2&=&e^{2\textrm{Im}\int_{\hat{r}(0)}^rd\hat{r}\hat{r}\log
\frac{\sqrt{\hat{r}}-\sqrt{2(M-\omega_k)}}{\sqrt{\hat{r}}-\sqrt{2M}}}\nn\\
&=&e^{4\pi[(M-\omega_k)^2-M^2]}=e^{\Delta S_{BH}}, \label{eq25}
 \eea
where $\Delta S_{BH}=S_{BH}(M-\omega_k)-S_{BH}(M)$ is the change
of the Bekenstein-Hawking entropy. So the emission rate is related
to the Bekenstein¨CHawking entropy, and the true radiation
spectrum of the black hole deviates from the strictly thermal one.
When considering the emission particle having a small energy, we
can also expand Eq.(\ref{eq25}) in terms of the energy $\omega_k$,
and reproduce the purely thermal spectrum.

Note that Eq.(\ref{eq25}) agrees with the underlying unitary
theory since the emission rate in quantum mechanics is expressible
as $\Gamma(i\rightarrow f)\propto |M_{fi}|^2$(phase space factor),
where $|M_{fi}|^2$ is the square of the amplitude for the emission
process, and the phase-space factor is given by summing over the
final states and averaging over the initial states. Since the
number of the initial/final states is the exponent of the
initial/final entropy, the emission rate is then given by
$\Gamma(i\rightarrow
f)\propto\frac{e^{S_{\textrm{final}}}}{e^{S_{\textrm{initial}}}}=e^{\Delta
S}$, which is consistent with our result (\ref{eq25}). Hence,
Eq.(\ref{eq25}) satisfies the underlying unitary theory and
provides a right correction to Hawking precisely thermal radiation
spectrum.

%%%%%%%%%%%%%%%%%%%%%%%%%%%%%%%%%%%%%%%%%%%%%%%%%%%%%%%%%%%%%
\section{Conclusion and Discussion}\label{sec3}
%%%%%%%%%%%%%%%%%%%%%%%%%%%%%%%%%%%%%%%%%%%%%%%%%%%%%%%%%%%%%
In this paper, we revisit the self-gravitation correction to black
hole radiation, and find that the precisely derived spectrum is
not only deviated from the purely thermal spectrum (already
present in Ref.\cite{r2}), but most importantly, is related to the
change of the Bekenstein-Hawking entropy and consistent with an
underlying unitary theory. The results are useful, as a starting
point, for the quantization of the gravity, and guarantees the
unitary evolution in black hole quantum radiation. Supposing the
emitted particle taking an small energy, we can also recover
Hawking purely thermal spectrum by expanding the result
(\ref{eq25}) to the first order.

Here, we have assumed that the mass of black hole is held fixed
whereas the total ADM mass of the hole-particle system are allowed
to fluctuate. Instead, we also fix the total ADM mass and allow
the black hole mass to fluctuate. In Ref.\cite{MF}, Parikh and
Wilczek adopted the latter assumption to develop this method, and
implemented Hawking radiation as a tunnelling process (later on,
many papers appear to discuss Hawking radiation via tunnelling
from different black holes\cite{t1,t2,t3,t4,t5,t6,t7}). In the two
cases, we can reproduce the same results, and provide a correction
to black hole radiation due to self-gravitation interaction. On
the other hand, the emission rate (\ref{eq25}) is semiclassically
derived by applying the WKB approximation and the saddle point
approximation. Such an approximation can only be valid in the low
energy regime. If we are to properly describe Hawking radiation,
then a better understanding of quantum gravity is a necessary
prerequisite.

\appendix

\section{The ADM decomposition}

The Hamilton description for general relativity demands a
(3+1)-dimensional decomposition for the 4-dimensional space-time.
When the arbitrary point $(t, x^i)$ on the 3-dimensional
space-like surface moves to the point $(t+dt, x^i)$ after the time
past $(t+dt)$, the surface would de deformed. Now, we can define
the deformation vector as $ \mathcal{N}^\mu\equiv \frac{\partial
X^\mu(t,x^i)}{\partial t}$, where $X^\mu(t,x^i)$ is the space-like
hypersurface. In $(t,x^i)$, the tangential vector defined by
$X_i^\mu\equiv \frac{\partial X^\mu}{\partial x^i}$ and the
longitudinal vector $n^\mu$ should satisfy \begin{equation}
 g_{\mu\nu}X_i^\mu n^\nu=0, ~~~~
g_{\mu\nu}X_i^\mu X_j^\nu=h_{ij},~~~~ g_{\mu\nu}n^\mu n^\nu =-1.
\label{eqA1}
 \end{equation}
The first term denotes the tangential vector and the longitudinal
vector are orthogonal each other. The induce metric $h_{ij}$ for
the arbitrary point on the space-like hypersurface is determined
by the second term. The third term implies the longitudinal vector
$n^\mu$ is time-like. The tangential vector $X_i^\mu$ and the
longitudinal vector $n^\mu$ constitutes the local tetrad on the
space-like hypersurface. Now making a (3+1)-dimensional
decomposition for the deformation vector on the local tetrad, we
have
\begin{equation}
\mathcal{N}^\mu=\mathcal{N}n^\mu+\mathcal{N}^iX_i^\mu,
\end{equation}
where $\mathcal{N}$ is called as the time shift, $\mathcal{N}_i$
as the space shift. For the general metric taking the form as
\begin{equation}
ds^2=g_{tt}dt^2+2g_{it}dx^idt+g_{ij}dx^idx^j.
\end{equation}
Noted that the deformation vector defined above means $\delta
X^\mu=\mathcal{N}^\mu dt$, which yields
\begin{equation}
g_{\mu\nu}\delta X^\mu \delta X^\nu=\mathcal{N}^\mu\mathcal{N}_\mu
dt^2=\big(\mathcal{N}^i
\mathcal{N}_i-\mathcal{N}^2\big)dt^2=g_{tt}dt^2,
\end{equation}
where $\mathcal{N}_i=h_{ij}\mathcal{N}^j$. Similarly, for the
component $g_{it}$, we have
\begin{equation}
g_{it}=g_{\mu\nu}X_i^\mu
\mathcal{N}^\nu=h_{ij}\mathcal{N}^j=\mathcal{N}_i.
\end{equation}
Now the action for the (3+1)-dimensional gravitational field can
be found by Eq.(\ref{eq2}).

\section{The Schwarzschild black hole in the Painlev\'{e} coordinate}

The Schwarzschild black hole in the Schwarzschild coordinate can
be written as
\begin{equation}
ds^2=-\Delta
dt_s^2+\Delta^{-1}dr^2+r^2\big(d\theta^2+\sin\theta^2d\phi^2\big),
\end{equation}
where $\Delta=\big(1-\frac{2M}{r}\big)$. Introducing the
Painlev\'{e} coordinate transformation as
\begin{equation}
dt_s=dt+\frac{\sqrt{1-\Delta}}{\Delta}, \label{B2}
\end{equation}
we have
\begin{equation} ds^2=-\Delta
dt^2+2\sqrt{1-\Delta}dtdr
+dr^2+r^2\big(d\theta^2+\sin\theta^2d\phi^2\big).
\end{equation}
The new metric has many attractive features. (i) The metric is
regular at the event horizon; (ii) The time coordinate $t$
registers the local proper time for radially free-falling
observers; (iii) In addition, it satisfies Landau's condition of
the coordinate clock synchronization. Integrating the Painlev\'{e}
coordinate transformation (\ref{B2}) yields
\begin{equation}
t_s=t-2\sqrt{2Mr}-2M\log\frac{\sqrt{r}-\sqrt{2M}}{\sqrt{r}+\sqrt{2M}}.
\end{equation}

\section*{Acknowledgments}
This work is supported by National Natural Science Foundation of
China with Grant Nos.10675051, 70571027, 10635020, a grant by the
Ministry of Education of China under Grant No.306022, and a
Graduate Innovation Foundation by Central China Normal University.

\end{document}